# Towards a framework for understanding societal and ethical implications of Artificial Intelligence

*Hacia un marco para comprender las implicaciones sociales y éticas de la Inteligencia Artificia*

Dr. V. Richard Benjamins
*Telefónica*
ORCID ID: 0000-0002-9412-7082

Dra. Idoia Salazar García
*Universidad CEU San Pablo*
ORCID ID: 0000-0002-9540-8740

**Resumen**

Artificial Intelligence (AI) is one of the most discussed technologies today. There are many innovative applications such as the diagnosis and treatment of cancer, customer experience, new business, education, contagious diseases propagation and optimization of the management of humanitarian catastrophes. However, with all those opportunities also comes great responsibility to ensure good and fair practice of AI. The objective of this paper is to identify the main societal and ethical challenges implied by a massive uptake of AI. We have surveyed the literature for the most common challenges and classified them in seven groups: 1) Non-desired effects, 2) Liability, 3) Unknown consequences, 4) Relation people-robots, 5) Concentration of power and wealth, 6) Intentional bad uses, and 7) AI for weapons and warfare. The challenges should be dealt with in different ways depending on their origin; some have technological solutions, while other require ethical, societal, or political answers. Depending on the origin, different stakeholders might need to act. Whatever the identified stakeholder, not treating those issues will lead to uncertainty and unforeseen consequences with potentially large negative societal impact, hurting especially the most vulnerable groups of societies. Technology is helping to take better decisions, and AI is promoting data-driven decisions in addition to experience- and intuition-based discussion, with many improvements happening. However, the negative side effects of this technology need to be well understood and acted upon before we launch them massively into the world.

**Abstract**

La inteligencia artificial (IA) es una de las tecnologías más discutidas en la actualidad. Existen muchas aplicaciones innovadoras como el diagnóstico y tratamiento del cáncer, la experiencia del cliente, los nuevos negocios, la educación, la propagación de enfermedades contagiosas y la optimización de la gestión de catástrofes humanitarias. Sin embargo, con todas esas oportunidades también viene una gran responsabilidad para garantizar una buena práctica justa de AI. El objetivo de este documento es identificar los principales desafíos sociales y éticos implicados por una aceptación masiva de la IA. Hemos examinado la literatura para los desafíos más comunes y los hemos clasificado en siete grupos: 1) Efectos no deseados, 2) Responsabilidad, 3) Consecuencias desconocidas, 4) Relaciones personas-robots, 5) Concentración de poder y riqueza, 6) Mal uso intencional, y 7) IA para armas y guerra. Los desafíos deben ser tratados de diferentes maneras dependiendo de su origen; algunos tienen soluciones tecnológicas, mientras que otros requieren respuestas éticas, sociales o políticas. Dependiendo del origen, diferentes actores podrían necesitar actuar. Cualquiera que sea la parte interesada identificada, el no tratar esos problemas dará lugar a incertidumbres y consecuencias imprevistas con un impacto social negativo potencialmente grande, afectando especialmente a los grupos más vulnerables de las sociedades. La tecnología está ayudando a tomar mejores decisiones, y AI está promoviendo decisiones basadas en datos, además de discusiones basadas en la experiencia y la intuición, con muchas mejoras en marcha. Sin embargo, los efectos secundarios negativos de esta tecnología deben entenderse bien y actuar antes de lanzarlos masivamente al mundo.

**Palabras clave:** Inteligencia Artificial, ética, robots, tecnología.

1. INTRODUCTION

Artificial Intelligence (AI) is on the rise. There are so many innovative applications of AI that everybody is speaking about it. It improves the diagnosis and treatment of cancer, improves customer experience, creates new business, improves education, predicts how contagious diseases propagate and optimizes the management of humanitarian catastrophes, to name just a few.

Artificial Intelligence already exists since the mid-fifties when John McCarthy first coined the term. Marvin Minsky's definition of AI was: "The science of making machines do things that would require intelligence if done by humans" (Dennis, 2016). AI has had its ups and downs previously (e.g. AI Winters). In a white paper published earlier by author of this paper Richard Benjamins (LUCA-Telefónica, 2016), was explained basic concepts of AI to provide some background information to better understand the many articles about AI appearing in the press and Internet. AI's current popularity is mostly due to the enormous progress in one of AI's subfields: Machine Learning (ML). There are three main reasons for this progress:

- The abundance of data. ML analyses data, and today there is so much more data available than decades ago.
- Increase in computational power. Moore's Law is still valid, and machines can process orders of magnitude faster and more than before.
- Deep Learning. An extended version of Neural Networks that, thanks to the two previous points, has increased enormously the performance of all kinds of classification and prediction tasks.

However, with all those opportunities also comes great responsibility to ensure good and fair practice of AI. We identified seven societal and ethical challenges for Artificial Intelligence that should be dealt with before AI is massively applied. Not treating those issues will lead to uncertainty and unforeseen consequences with potentially large negative societal impact. It is therefore no surprise that many governments currently have set up national initiatives to discuss many of those issues (e.g. in the UK (Committee on AI, 2018) and France (Le Gouvernement Republique Française, 2017).

2. OBJECTIVES

The study and understanding of the societal and ethical aspects of Artificial Intelligence is a new, yet fast-growing area, where not much scientific literature exists. However, many informal articles and publications discuss the topic on the Internet, in the press and as outcomes of closed workshops with experts. The objective of this paper is to help shape the area by proposing a framework to classify the different types of societal and ethical challenges such that each can be properly studied and addressed through research and experiments.

**Specific objectives:**

1. Analyse the social impact that the development of Artificial Intelligence is having, through the study of current real cases.
2. Identify the advantages and disadvantages of the massive implementation of this technology.
3. Identify old ethical dilemmas and compare them with the current ones derived from the implementation of these technologies.
4. Offer a theoretical frame of reference for further studies related to the social impact of Artificial Intelligence.
5. Determine future prospects, in the medium term, of this technology, once the seven major challenges are analysed.

3. METHODOLOGY

The study and understanding of the societal and ethical aspects of Artificial Intelligence is a new, yet fast-growing, area, where not much scientific literature exists. However, many white papers from different governments and organizations discuss the topic, knowing its relevance and high impact in the next future. In this paper we have analyse several white papers, defining white paper as: authoritative report or guide, made up by high qualified experts, that informs readers concisely about a complex issue and presents the issuing body's philosophy on the matter (Gordon and Graham, 2003). We also analyse different reports published in some of the most worldwide prestigious magazines specialized in technology as Wired.

We also have consulted some scientific articles from databases accessible through the Internet, as Scopus bibliographic database which supports the quality of these publications (De Granda-Orive et al., 2013). This is the largest database of abstracts and literature reviewed by peers and has intelligent tools that allow to control, analyse and visualize academic research. Given its wide coverage, both geographical and thematic, it is considered ideal to be used for bibliographic reviews (Codina, 2018). The main search criteria have been delimited by keywords including Boolean inclusion (AND) or exclusion (NOT) operators. Date range filters have also been used. Likewise, these same search criteria have been applied in the database Web of Science, where the references of the main scientific publications of any discipline of knowledge, both scientific and technological, humanistic and sociological since 1945 are detailed.

In addition, it has been of great help for the acquisition of the necessary knowledge for this research, the collaboration of the authors of this article in the online forum International Ethically Aligned Design (EAD), in which professionals of international relevance expose their point of view on how to create the best ethical code of Artificial Intelligence, taking into account the social impact of this technology. As well as in the discussion forums of the European AI Alliance, a body dependent on the European Commission, of which the authors are also a member.

Also, the knowledge of the authors necessary for the realization of this paper derives from several investigations related to other areas of the social impact of Artificial Intelligence, acquired as researchers of the SIMPAIR group (Social Impact of Artificial Intelligence and Robotics).

4. ANALYSIS AND RESULTS OF THE INVESTIGATION

4.1. Seven societal and ethical challenges for Data and AI

The challenges that AI, and therefore Data as well, face include non-desired side effects, liability questions, yet unknown consequences, the relation human-robots, increasing concentration of power and wealth, intentional bad uses, and AI for weapons and warfare. Below, we explain each of them.

1. *Non-desired side effects*

a)  While Machine Learning is able to solve complex tasks with high performance, it might use information that from a society or humanity perspective is undesired. For example, deciding whether to provide a loan to people based on race or religion is not something our societies accept. While it is possible to remove those "undesired" attributes from data sets, there are other less obvious attributes that highly correlate with those "undesired" attributes whose removal is less straightforward. Machine Learning is objective and finds whatever relation there is in the data regardless of specific norms and values.

b)  A related issue is so-called bias of data sets. Machine Learning bases its conclusions on data. However, the data itself might be biased by not being representative for the group of people to which the results are applied. For instance, finding trends on school performance using mostly white schools will not provide insights applicable to all schools. Research has shown that ML takes over any bias from humans (Burgess, 2017). What we need is so-called "Fair" Machine Learning, addressing the issues in point a) and b). (World Economic Forum, 2018).

c)  Apart from bias in the training data, bias can also come from the algorithm. A Machine Learning algorithm tries to be as accurate as possible when fitting the model to the training data. Accuracy can be defined in terms of so-called "false positives" and "false negatives", often through a so-called confusion matrix. But the definition of this "accuracy" measure, whether it tries to optimize only false positives or only false negatives, or both, has an important impact on the outcome of the algorithm, and therefore on the groups of people affected by the AI program. In safety-critical domains such as health, justice, and transport defining "accuracy" is not a technical decision, but a domain or even a political decision.

d)  Deep learning algorithms can be highly successful but have a hard time to explain why they have come to a decision. For some applications, the explanation of decisions is an essential part of the decision itself, and lack of that makes the decision unacceptable. For example, a "robo-judge" deciding on a dispute between a customer and a health insurer is unacceptable without the explanation of the decision. This is referred to as the "Interpretability" problem. The book "Weapons of math destruction" gives many interesting examples of this.

e)  Data privacy, transparency and control. All data and AI system exploit data, and many times this is personal data. Using all this personal data has as side effect that privacy may be compromised, even if it is unintentionally. The recent scandal of Cambridge Analytica / Facebook shows that this is a bigger issue than we might have thought.

To avoid those effects, people sometimes refer to the need for FATE AI (Fair, Accountable, Transparent and Explainable Artificial Intelligence).

*2. Liability.* When systems become autonomous and self-learning, accountability of behaviour and actions of those systems becomes less obvious. In the pre-AI world, incorrect usage of a device is the accountability of the user, while device failure is accountability of the manufacturer. When systems become autonomous and learn over time, some behaviour might not be foreseen by the manufacturer. It is therefore unclear who would be liable in case something goes wrong. A clear example of this are driverless cars. Discussions are ongoing whether a new legal person needs to be introduced for self-learning, autonomous systems, such as a legal status for robots, but it is generating some controversy.

*3. Unknown consequences of AI.* The positive aspects of AI may have some consequences of which we don't know yet how they will work out.

a) AI can take over many boring, repetitive or dangerous tasks. But if this happens at a massive scale, maybe many jobs might disappear, and unemployment will skyrocket?

b) If less and less people work, then the government will receive less income tax, while costs of social benefits will increase due to increased unemployment. How can this be made sustainable? Should there be a "robot tax"? How to be able to pay pensions when increasingly less people work?

c) Is there a need for a universal basic income (UBI) for everybody? If AI takes most of the current jobs, what do all unemployed people then live from?

*4. How should people relate to robots?* If Robots become more autonomous and learn during their "lifetime", then what should be the (allowed) relationship between robots and people? Could one's boss be a robot, or AI system? In Asia, robots are already taking care of elderly people, accompanying them in their loneliness. And, could people get married to a robot?

An initial overview of possible interactions between humans and robots includes (Dautenhahn, 2017):

a) Long-term interactions: in which robots cohabitate with humans in their homes, and work places.

b) Robots in therapy, rehabilitation and supporting the elderly: Assistive robotics is a growing application domain for service robots. It involves

    critical safety and ethical issues, for example when robots take the role of assisting vulnerable people or people with special needs. It is a fact that today, approximately 10 percent of the world's population is over 60 years old and by 2050 this proportion will have more than doubled. We need to incorporate artificial intelligence techniques to support older adults and help them cope with the changes of aging, in particular with cognitive decline (Pollack, 2005)

c) Multimodal interactions, expressiveness, and conversational skills in interactions: Research aiming at providing robots with human-like features and qualities is expanding. We are trying to create robots with a similar appearance to us. And not only this. We are also trying to provide "their faces" with natural expressions that normally would only correspond to humans.

d) Social learning and skill acquisition via teaching and imitation: This theme involves research on robots that can adapt to changing environments and requirements, that can 'grow' with increasing levels of skills and knowledge they acquire, and that can be programmed indirectly by demonstrating tasks. Once acquired, the robot could also transfer the newly learnt skills to other robots. If this turns out a real possibility, it could lead to an ethical problem.

e) Cooperation and collaboration in human-robot teams: Robots and humans will not only live side by side, but they also need to work hand in hand, each one of them performing specific tasks, but helping each other. This interaction, mostly at the beginning, might be quite difficult taking into account the concept of 'machine', and all its implications, that humanity has nowadays.

f) Detecting and understanding human activity: This will be also a problem in the near future for this interaction. Many times, humans do things for reasons that could be out of logic for a machine.

These are just some of the possible interactions in the relationship between human and robots. Each one of them will probably need further study to get to relevant conclusions.

    5. *Concentration of power and wealth in a few very large companies.* Currently, AI is dominated by a few large digital companies, including GAFAM (Google, Amazon, Facebook, Apple, Microsoft) and some Chinese mega companies (BAT - Baidu, Alibaba and Tencent). This is mostly due to those companies having access to massive amounts of propriety data. This might lead to an oligopoly. Apart from the lack of competition, there is a danger that those companies

keep AI as proprietary knowledge, not sharing anything with the larger society other than for the highest price possible. Another concern of this concentration is that those companies can offer high-quality AI as a service, based on their data and propriety algorithms (black box). When those AI services are used for public services, the fact that it is a black box (bias, undesired attributes, performance, etc.), raises serious concerns, like when the LA Police Department announced that it uses Amazon's face recognition solution (Rekognition) for policing. The Open Data Institute in London has started an interesting debate on whether AI algorithms and Data should be closed, shared or open.

6. Intentional bad uses. All the points mentioned above are issues because AI and Data are applied with the intention to improve or optimize our lives. However, like any technology, AI and Data can also be used with bad intentions[1]. Think of AI-based cyberattacks[2], terrorism, influencing important events with fake news[3], etc.

7. The last challenge is related to the application of *AI for warfare and weapons*, especially for lethal autonomous weapons systems (LAWS). This usually implies an explicit, political decision, that some will consider "good use", while other might call it bad use of AI. Some organizations are working on an international treaty to ban "killer robots". The issue recently attracted attention due to Google employees sending a letter to their CEO questioning Google's participation in defence projects.

### 4.2. Stakeholders and potential actions

While there is ample debate ongoing about many of those issues (e.g. AI and the future of work), today it is unclear what solutions there will be for some of the challenges. What we can define, however, are relevant actions to work on as an approach for dealing with those challenges. Each of the actions might involve different stakeholders. While executing the approach, because of the uncertainty involved, adaptions need to be made in a learning by doing process.

- Governments and institutions need to think about strategies and approaches to identify the issues along with their solution directions. The GDPR is a small, but important step into that direction. Several national governments are already working on this through multidisciplinary committees of experts (Committee on AI, 2018, Le Gouvernement Republique Française, 2017). Maybe governments should ensure the availability of rich and sufficiently varied open datasets to minimize unfair bias.
- Private enterprises need to start thinking about self-regulation and about where they stand. They should be clear on how responsible they want to act and become.
- Probably a one-size-fits-all approach will not work, as some AI and Big Data applications have less potential negative side effect than others. E.g., decisions in marketing have probably less negative side effects than decisions about insurance premiums or medical diagnosis. Decisions made for so-called "safety-critical" systems may need to be validated and verified by formal mathematical procedures to ensure their correct functioning under all possible circumstances.
- GDPR (Europe's General Data Protection Regulation) is a step forward regarding protection of personal data. A clear distinction needs to be made between Data & AI applications using personal data versus those using aggregated, anonymized data. Applications using personal data will need an explicit and transparent user consent (as provisioned in the GDPR). This is not needed when aggregated, anonymized data is used, but, as a matter of transparency, one might argue for the "right to be informed" when users' (aggregated, anonymized) data is used for applications.
- AI and Data should not only be used for commercial opportunities, but also for social opportunities, such as Data for Good and AI for Good, initiatives whose aims are to support achieving the United Nations Sustainable Development Goals (SDGs).
- The Open Data approach should be extended to Open Algorithms, to enable the benefits of private data, while not increasing the privacy and commercial risks of private enterprises. In this approach, algorithms are sent to the data, so data remains in its original premises, rather than the usual other way around, where data is stored centrally and then algorithms are run on the data.
- Code of conducts should be in place for all professional families that contribute to Data and AI applications. This should reduce the likelihood of bad uses and unintended negative side effects.
- International, multidisciplinary committees should be put in place to oversee and monitor the uses of Data and AI across the world and raise alerts when needed. Something like the Civil Aviation Authority for airplane crashes.

---

[1] (2018, 02). The Malicious Use of Artificial Intelligence: Forecasting, Prevention, and Mitigation. Open Data Institute. 05, 2018, *https://img1.wsimg.com/blobby/go/3d82daa4-97fe-4096-9c6b-376b92c619de/downloads/1c6q2kc4v_50335.pdf*

[2] (2018, 02). Artificial intelligence poses risks of misuse by hackers, researchers say. Reuters. 05, 2018, *https://www.reuters.com/article/us-cyber-tech/artificial-intelligence-poses-risks-of-misuse-by-hackers-researchers-say-idUSKCN1G503V*

[3] (2018, 04). Artificial intelligence is making fake news worse. Business Insiders. 05, 2018, *http://uk.businessinsider.com/artificial-intelligence-is-making-fake-news-worse-2018-4*

The following table relates the identified concerns with potential actions of relevant stakeholders.

| Challenge category | Concern | Stakeholder to act | Type of action | Potential action |
|---|---|---|---|---|
| Non-desired side effects | Non-desired side effects | Companies and organizations | Policy | Code of conducts to raise awareness of non-desired consequences to designers, developers and users of AI |
| | Undesired attributes | Companies and organizations | Technical | Find correlations between sensitive data and other variables |
| | Bias | Companies and organizations | Technical | Develop or use tool that checks bias of data set w.r.t. target group |
| | | Governments | Policy | Publish bias-free Open Data sets |
| | Interpretability | Companies and organizations | Technical | Perform and use research on explainable algorithms |
| | | Governments | Ethical | Discussions on in what areas AI needs to be explainable |
| | Privacy | Companies and organizations | Policy | Privacy by design, and transparent privacy policies |
| | | Companies and organizations | Technical | Send algorithms to data rather than transferring data out of organizations increasing data breach risk |
| Liability | Liability | Governments | Policy | Discussion on liability of self-learning, autonomous systems |
| Unknown consequences of AI | Work and jobs | Companies and organizations | Policy | Training and up-skilling of employees to the digital world |
| | | Governments | Policy | Adapt educational system for AI future. |
| | | Governments and institutions | Policy | Evaluate impact of Universal Basic Income and sustainability of welfare system |
| Relation people-robots | Relation people-robots | Governments | Ethical | Study the possible ways in which people can relate to robots, from collaboration to medical assistance to marriage |
| Data & wealth concentration | Data concentration | Governments | Policy | Evaluate what can be done to reduce dependency on a few technological giants with disproportional amount of data |
| Intentional bad uses | Intentional bad uses | Governments | Policy | Strengthen cybersecurity units with AI and data expertise |
| AI and warfare | AI and warfare | Governments | Ethical | Discuss explicitly how far we should go with so-called "killer robots" |
| General world-wide impact of AI | General world-wide impact of AI | Governments | Policy | Create international organizations with authority over important AI developments and projects in the world |
| | | Companies, organizations and Governments | Technical | Use AI for social purposes to help achieve the UN's Sustainable Development Goals. |

*Source:* Prepared by the authors

5. CONCLUSIONS

With AI become increasingly more widespread, it is important that we are prepared to ensure that our societies and economies continue to function well with the expected massive uptake of Data & AI applications. In this paper, we have identified seven types of challenges to be dealt with before massively using AI and Big Data in our societies. We then looked at what type of actions are required by the different stakeholders to maximize the likelihood of the issues to be tackled. While some of the required actions are already in progress, several others are only starting. In general, we expect that significant debate is needed to take final decisions and there is no guarantee that different nations in the world will come to the same conclusions.

However, while much work still needs to be done, we should neither forget that today, without massive uptake of AI, we don't live, and have never lived, in an ideal world. Think of the large number of humans that have taken, are taking and will take extremely wrong decisions, with hugely negative consequences for humanity. People are "human" and therefore decisions inevitably are biased by personal experiences and opinions. And many decision makers have taken –with good intent– important measures that have had serious negative side effects. So, while there are risks associated with the massive uptake of AI & Data, there is probably more to win than to lose with those technologies. Moreover, humanity has ample experience in how to manage or recover from negative consequences of (wrong) decisions.

If there is a final deadline before we would need to have solved the seven challenges, it will probably be the Singularity, the point in time when Artificial Intelligence will lead to machines that are smarter than human beings. However, even if this point may never arrive, it is good that societies start discussing the issues now, through a common vocabulary and framework, rather than waiting until it is (too) late.

REFERENCES

Burgess, M (2017): Just like humans, artificial intelligence can be sexist and racist. *Wired Magazine*. [Disponible en: *https://www.wired.co.uk/article/machine-learning-bias-prejudice*] [Consultado el 01/08/2018]

Committee on AI (2018): AI in the UK: ready, willing and able? *Authority of the House of Lords*. [Disponible en: *https://publications.parliament.uk/pa/ld201719/ldselect/ldai/100/100.pdf*] [Consultado el 10/08/2018]

Dautenhahn, K (2017): Methodology & Themes of Human-Robot Interaction: *International Journal of Advanced Robotic Systems*. [Disponible en: *http://journals.sagepub.com/doi/full/10.5772/5702*] [Consultado el 04/10/2018]

Dennis, M (2016): Marvin Minsky. American Scientist. *Encyclopedia Britannica*. [Disponible en: *https://www.britannica.com/biography/Marvin-Lee-Minsky*] [Consultado el 03/07/2018]

Le Gouvernement Republique Française (2017): Rapport de synthèse France Inteligence Artificielle. [Disponible en: *https://www.economie.gouv.fr/files/files/PDF/2017/Rapport_synthese_France_IA_.pdf*] [Consultado el 10/08/2018]

LUCA-TELEFÓNICA (2016): White Paper: Surviving the AI Hype-Fundamental concepts to understand Artificial Intelligence. [Disponible en: *https://www.slideshare.net/LUCA-D3/surviving-the-ai-hype-fundamental-concepts-to-understand-artificial-intelligence-70400058*] [Consultado el 06/09/2018]

Pollack, M (2005): Intelligent Technology for an Aging Population: The Use of AI to Assist Elders with Cognitive Impairment. [Disponible en: *https://www.aaai.org/ojs/index.php/aimagazine/article/view/1810tps://www.slideshare.net/LUCA-D3/surviving-the-ai-hype-fundamental-concepts-to-understand-artificial-intelligence-70400058*] [Consultado el 06/09/2018]

World Economic Forum (2018): White Paper: How to Prevent Discriminatory Outcomes in Machine Learning. *Global Future Council on Human Rights*. [Disponible en: *http://www3.weforum.org/docs/WEF_40065_White_Paper_How_to_Prevent_Discriminatory_Outcomes_in_Machine_Learning.pdf*] [Consultado el 04/08/2018]